# Phonon modes contribution in thermal rectification in graphene-C₃B junction: A molecular dynamics study

Leila Kiani [a], Javad Hasanzadeh [a*], Farrokh Yousefi [b]

[a] Department of Physics, Takestan Branch, Islamic Azad University, Takestan, Iran

[b] Department of Physics, University of Zanjan, Zanjan, Iran

**Abstract:**

In this work, using non-equilibrium molecular dynamics simulation, we implement a series of simulation under positive and negative temperature gradient in order to investigate the thermal rectification in the graphene-C₃B junction (GCB). The dependence of thermal rectification on temperature difference between the hot and the cold baths is obtained. The important quantity that we present here, is the in-planes and out-of-plane phonon modes contribution in the thermal rectification. We see that the Y mode has high and positive thermal rectification while that the X and Z modes have small and negative thermal rectifications. Thermal rectification for Y mode increases sharply beyond $\Delta T > 30K$ to ~150% but for X and Z modes decrease slowly by increasing $\Delta T$. Our results show that Y mode has major role in the thermal rectification. Moreover, the underlying mechanisms that leads to the thermal rectification and also Kapitza resistance at the interface are studied via the phonon density of states (DOS).

---

[*]Corresponding author: j.hasanzadeh@yahoo.com

**Introduction**

Graphene [1], which is a two dimensional (2D) monolayer with honeycomb form, has attracted many people from around the world due to excellent mechanical [2–4], high thermal conductivity [5–8], optical [9] and electronic properties [10]. Although graphene has amazing properties, there are some limitations in real application. For example, graphene has zero-band-gap electrical property, therefore we have restriction to use it in the semiconductor industry and nanoelectronic devices. Therefore, many experimental researches were done to find new 2D materials such as $MoS_2$ [11], h-BN [12], graphane [13,14], $C_3N$ [15], $C_3B$ [16,17], borophene [18], phosphorene [19] and dozens of others for different goals. Thermal properties of these 2D materials were studied very well in the last decade [20–22]. Moreover, it is interesting to determine the thermal properties such as thermal rectification (thermal diode) and thermal logic gate from the hybrid of these materials. Compared with electrical diode, thermal rectifier allows heat current to pass in one direction more than opposite direction under positive and negative temperature gradient. Thermal rectifiers usually can be constructed of two or more materials (hybrid) or in the mass graded systems [23–27]. A lot of investigation were implemented over thermal rectification using hybrid systems with graphene. It should be noted that the thermal rectification depends significantly on the interface type and also used materials. Rajabpour et al. [25] have studied thermal rectification in hybrid graphene-graphane. The have found that the thermal rectification is insensitive on the sample length in the range of (20nm-100nm) and equals to ~20%. Shavikloo et al. [28] have also examined thermal rectification in the partially hydrogenated graphene with grain boundary type 5-7-5-7. They have shown that the thermal rectification changes with mean temperature of the system and decreases with increasing the difference temperature between the hot and the cold baths ($\Delta T$). In the other hybrid system, graphene connected to $C_3N$ was studied by Farzadian et al. [29]. The important parameter that was examined on thermal rectification is $\Delta T$. In the last system, they have found that there is a thermal resistance (Kapitza resistance) at the interface between two sides. The Kapitza resistance in a thermal rectifier depends on the temperature gradient direction. In the graphene-$C_3N$ the Kapitza resistance are $5.67 \times 10^{-11}$ and $5.02 \times 10^{-11} m^2 K/W$ for two opposite temperature gradients. Therefore, the heat current prefers one direction than another. Moreover, nonporous graphene which was synthetized recently [22] was investigated in conjunction with pristine graphene. The system shows a small thermal rectification about 4.66% and 6.01% for armchair and zigzag directions, respectively.

It is noteworthy that Mortazavi et al. [30,31] employed machine-learning method for interatomic interactions in order to obtain thermal conductivity with more accuracy.

In this investigation, we study thermal properties and thermal rectification in the hybrid graphene-$C_3B$. The effect of the $\Delta T$ is specially examined on the in-planes and out-of-plane phonon modes thermal rectification. Moreover, we calculate the Kapitza resistance at the interface and also phonon density of states (DOS) to show phonon scattering in the graphene-$C_3B$ junction.

## Computational method

In this work, we implemented a series of non-equilibrium molecular dynamics simulation to investigate thermal rectification in the G-C3B monolayer using the Large-scale Atomic/Molecular Massively Parallel Simulator (LAMMPS) [32] as a classical MD package. Here, the simulated system was constructed from graphene and $C_3B$ (see Fig. 1). The hybrid system is a sheet with width of 10 nm and length of 20 nm.

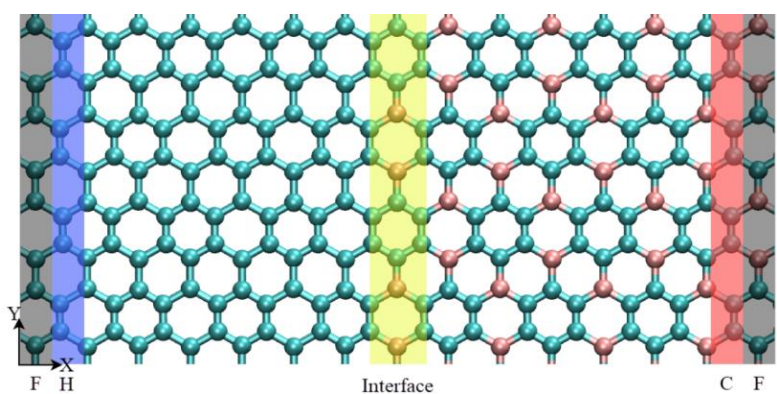

FIG. 1. Schematic of the simulated system. Left and right sides are graphene and $C_3B$. F, H and C indicate the fixed, hot and cold regions.

The periodic and free boundary conditions were applied to in-plane and out-of-plane directions, respectively. To describe the interaction potential between atoms in the structure, we used the three-body Tersoff potential [33,34]. This potential is appropriate for covalent systems and can model very well creating and breaking bonds between atoms. The timestep was considered 1 fs. The velocity Verlet algorithm was performed to integrate the equation of motion. First, to eliminate the extra stress in the system and relax it, we implemented NPT simulation with Nose-Hoover [35] thermostat and barostat at room temperature and zero pressure. Because of extra stress can affect the heat current we remove it.

After that, as shown in Fig. 1, we chose two slabs with width 1 nm at two ends of the G-C3B monolayer as fixed region. The atoms inside the fixed regions remains motionless during the simulation time after relaxing the system. The two other slabs near the fixed slabs were considered as the hot and the cold baths. We applied NVT ensemble (with Nose-Hoover thermostat) to the thermal baths to keep the temperature of them on a specified value. The difference temperature between the two thermal baths is $\Delta T$. Therefore, the temperature $T + \frac{\Delta T}{2}$ and $T - \frac{\Delta T}{2}$ were considered for the hot and the cold baths, respectively, which $T$ is the mean temperature of the system. For the region between two baths (middle region), we assumed the NVE ensemble. In the described situation, a temperature gradient was established along the system. To calculate the temperature gradient, we divided the whole of the system along the $X$ direction to 20 slabs. Then, the temperature inside each slab was calculated using the equation below,

$$T = \frac{1}{3Nk_B}\sum_{i=1}^{N} m_i v_i^2 \qquad (1)$$

where $N$ and $k_B$ are the number of atoms within a specified slab and Boltzmann constant, respectively. Also, $m$ and $v$ are atomic mass and velocity. The slope of the temperature profile represents the temperature gradient along the system. Moreover, to obtain the heat flux along temperature gradient direction, we let the system to reach steady-state during 3 ns simulation time. After that, we calculated three phonon modes contribution to the heat flux. The total heat flux $J$ is equal to the summation of the phonon modes contribution as shown below,

$$J = J_x + J_y + J_z \qquad (2)$$

where $J_x$, $J_y$ and $J_z$ are the heat fluxes along $X$ direction that the first two are generated by in-planes phonon modes ($x$, $y$) and the third one is produced by the out-of-plane phonon mode ($z$). The heat fluxes are represented as below [36],

$$J_x = \frac{1}{V}\Sigma_i(e_i v_{ix} - S_{i,xx}v_{ix}) \tag{3.a}$$

$$J_y = -\frac{1}{V}\Sigma_i S_{i,xy}v_{iy} \tag{3.b}$$

$$J_z = -\frac{1}{V}\Sigma_i S_{i,xz}v_{iz} \tag{3.c}$$

where $i$ runs over all atoms in the middle region. $V$ and $v$ are the middle volume and atomic velocity, respectively. Also, $S_{ab}$ is the per-atom stress tensor element. The stress tensor elements are calculated using the equation below [37],

$$S_{ab} = -[mv_a v_b + \frac{1}{2}\Sigma_{n=1}^{N_p} r_{1a}F_{1b} + r_{2a}F_{2b}] \tag{4}$$

resulting 3 components of the stress tensor. In Eq. (4), $a$ and $b$ represents $x$, $y$, and $z$. The first term indicates the contribution of the kinetic energy for atom $i$. The second term shows pairwise energy contribution. In this equation $r_1$ and $r_2$ are the position of the two atoms in the pairwise interaction. Also the $F_1$ and $F_2$ are the forces applied on the two atoms originating the pairwise interaction.

To calculate the thermal rectification factor, we considered two cases: (a) positive and (b) negative gradient. By definition, in the positive gradient, we assumed the hot region on the left side and the cold region on the right side and vice versa for the negative gradient (see Fig. 1). In both cases, we calculated the heat flux (for three phonon modes and total) and then obtained the total thermal rectification factor via the equation below [22],

$$TR(\%) = \frac{J_{l\rightarrow r} - J_{r\rightarrow l}}{J_{r\rightarrow l}} \times 100 \tag{5}$$

which $J_{l \to r}$ ($J_{r \to l}$) is the heat flux passes from left to right (right to left).

To examine the underlying mechanisms of thermal rectification, we need calculate the phonon density of states (DOS). The DOS was obtained through calculating Fourier's transformation of the velocity autocorrelation (represented by <>) using the equation below [38],

$$DOS(\omega) = \sum_l \frac{m_l}{k_B T} \int_0^\infty e^{-i\omega t} <\vec{v}(0).\vec{v}(t)> dt \qquad (6)$$

where $\omega$ is the angular frequency. The summation is over all atom types. Also $m$, $k_B$, $T$, $\vec{v}$ are the atomic mass, Boltzmann constant, temperature, and velocity vector, respectively.

**Results and discussion**

All of the NEMD simulations were carried out to extract the thermal rectification in the hybrid G-C3B monolayer. As the first result, we calculated the temperature profile along the $X$ direction in a sheet with a length of 20 nm. The temperature of each slab was calculated according Eq. (1) and averaged over 3 nm ($3 \times 10^6$ steps).

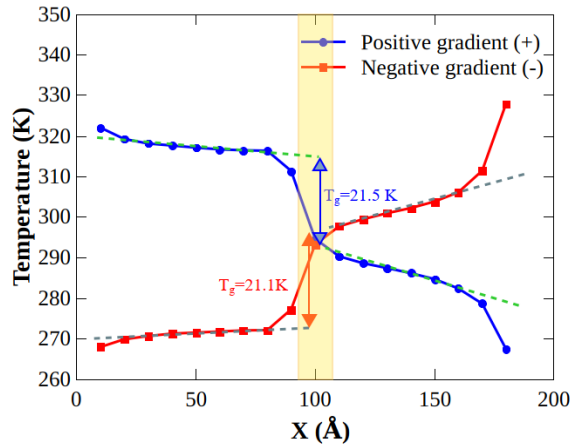

FIG. 2. The temperature profile along the G-C3B monolayer. The blue and red-colored show the temperature profile for positive and negative gradients, respectively. The $T_g$ is the temperature gap at the interface. Here $\Delta T = 70\ K$.

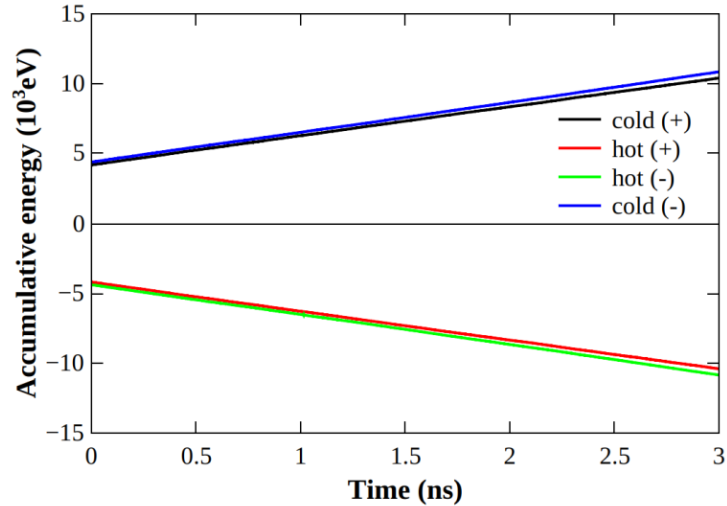

FIG. 3. The accumulative energy that inserted to the hot and extracted from the cold versus time. The slope of the curves is the heat current. According to these curves, the heat current which inserted or extracted from baths are the same.

As shown in Fig. 2, there is a temperature gap at the interface. The temperature gap appears when two different media are connected or if there is a boundary between the two same materials. This issue indicates that there is a thermal resistance (known as Kapitza resistance [39]) at the interface. The Kapitza resistance was calculated through $R_K = \frac{T_g}{J}$, where $T_g$ is the temperature gap at the interface. The Kapitza resistance allows less heat current to pass through the interface. After calculation the heat flux, the Kapitza resistance will be obtained.

Moreover, we checked the steady-state heat current in the system. For this purpose, we obtained the accumulative energy that added to the hot region and removed from the cold region during simulation time. The slope of the accumulative energy is the heat current. As represented in Fig. 3, the heat current (energy per time) that added to the hot or removed from the cold are the same for both of positive and negative gradients. Therefore, the system reaches the steady-state.

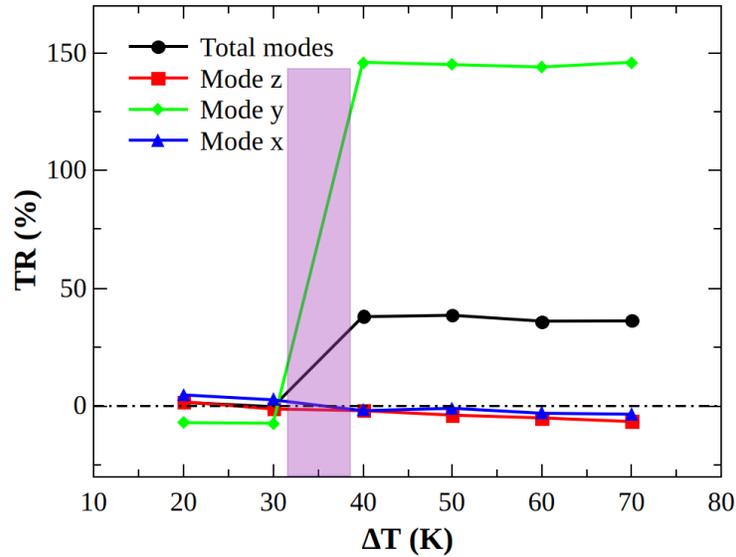

FIG. 4. The thermal rectification versus temperature difference between two baths. Contribution of the in-planes and out-of-plane in rectification were explored.

The Kapitza resistances for positive gradient (left to right) and negative gradient (right to left) were obtained $5.4 \times 10^{-12} m^2 K/W$ and $7.3 \times 10^{-12} m^2 K/W$, respectively. The Kapitza resistance values show that the total heat current prefers to pass to the right at $\Delta T = 70\ K$.

In this paper, we explored the thermal rectification in the G-C3B. It is a fundamental study to understand the contribution of the in-planes and out-of-plane phonon modes in the thermal rectification. According to Fig. 4, the total thermal rectification before 30 K is zero approximately while that for 30 K up to 70 K, we have ~37% thermal rectification. This behavior shows that the hybrid G-C3B can be promising in practical usage due to its constant value of thermal rectification. The reason of the curve behavior is that when the temperature difference ($\Delta T$) increases, two factors can influence the thermal rectification. First, by increasing $\Delta T$ phonons with high energy (high frequency) will be excited that leads to an increase in the heat current (positive factor). Second, by increasing $\Delta T$ the temperature of the hot side of the G-C3B increases. Therefore, the phonon-phonon scattering will be increase. This subject arises a thermal resistance and decreases the heat current (negative factor). These factors compete with each other and determine the heat current value. Here, our results show that in the total rectification for $\Delta T > 30\ K$, the $J_{l \to r}$ is larger than $J_{r \to l}$ according to Eq. 5.

For calculation thermal rectification of two phonon modes *X* and *Z*, we used $TR_m(\%) = \frac{(J_{l \to r})_m - (J_{r \to l})_m}{(J_{l \to r})_m} \times 100$, where m is *X* or *Z*, and for phonon mode *Y*, we employed $TR_Y(\%) = \frac{(J_{l \to r})_Y - (J_{r \to l})_Y}{(J_{r \to l})_Y} \times 100$. A remarkable point is that phonon modes *X* and *Z* have small and negative thermal rectification (-1% ~ -4%) while that for mode *Y* the thermal rectification is large and positive (~145%). This means that the mode *Y* has a great impact on the total thermal rectification. It should be noted that the summation thermal rectification of three phonon modes is not equal to total rectification because thermal rectification for each mode was calculated separately. Therefore, we have,

$$TR_{total} \neq TR_X + TR_Y + TR_Z \qquad (7)$$

Another significant point is that for $\Delta T \leq 30\ K$ the thermal rectification for modes *X* and *Z* is positive while that for mode *Y* is negative. The competition between two positive and negative factors determines the behavior of the curves.

Moreover, to understand the underlying mechanisms of the thermal rectification, we calculated the DOS on both sides of the interface according to Eq. 6. As shown in Figs. 5 and 6, the total and for the three modes DOS were calculated. To calculate DOS, tow narrow layers with width 3 nm on both sides and the near interface was chosen. The mismatching between two DOSs shows that when phonons move from the hot side to the cold side scatter at the interface. Therefore, the phonon spectra change from one side to another.

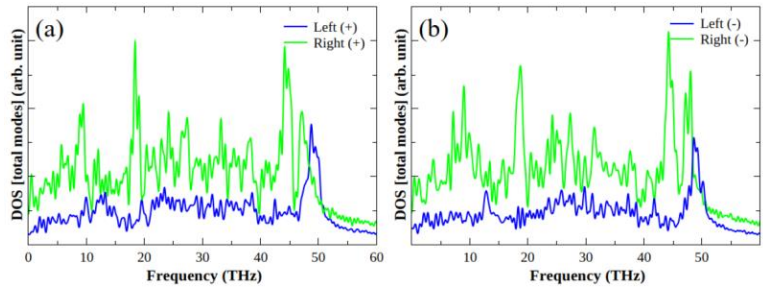

FIG. 5. The phonon density of states for total modes on both sides of the interface. (a) For positive and (b) negative gradient. The mismatching between two green and blue-colored indicates phonon scattering at the interface.

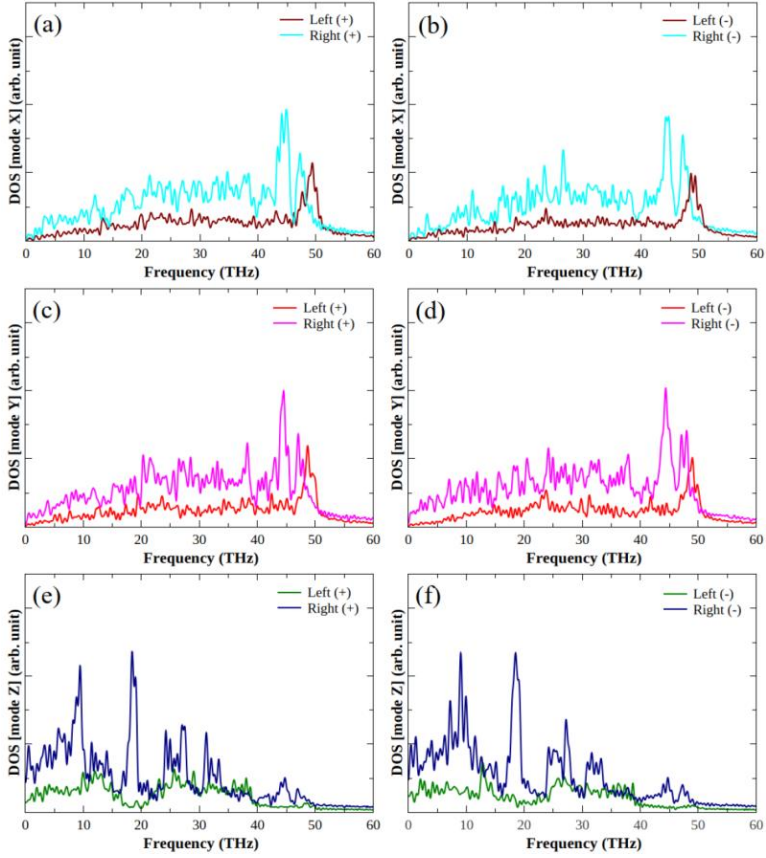

FIG. 6. The DOS for three in-planes and out-of-planes modes. (a) and (b) are for mode *X*, (c) and (d) for mode *Y*, (e) and (f) for mode *Z*.

**Conclusion**

In the present work, the non-equilibrium MD simulation were implemented to explore thermal rectification of in-planes (*X* and *Y*) and out-of-plane (*Z*) phonon modes as well as the total modes. We also investigated the temperature difference ($\Delta T$) dependence of the thermal rectification. According to the obtained results, total thermal rectification is about ~37% for $\Delta T > 30\ K$ and zero for $\Delta T \leq 30\ K$. We found that thermal rectification for mode Y is positive and significantly large (~145%) than other modes X and Z (-1% ~ -4%). This means that the mode Y has a great influence on the total thermal rectification while that modes X and Z have small and ignorable rectification. Moreover, we calculated the phonon density of states for different phonon modes to understand underlying mechanisms of thermal rectification. Phonons scattered at the interface that leads to thermal resistance there. The thermal resistance depends on the temperature gradient direction. Therefore, the heat current prefers one direction (graphene to $C_3B$) to another.